# Additive Update Algorithm for Nonnegative Matrix Factorization

Tran Dang Hien
Vietnam National University
hientd_68@yahoo.com

Do Van Tuan
Hanoi College of Commerce and Tourism
Hanoi – Vietnam
dvtuanest@gmail.com

Pham Van At
Hanoi University of Communications
and Transport
phamvanat83@vnn.vn

*Abstract*—**Nonnegative matrix factorization (NMF) is an emerging technique with a wide spectrum of potential applications in data analysis. Mathematically, NMF can be formulated as a minimization problem with nonnegative constraints. This problem is currently attracting much attention from researchers for theoretical reasons and for potential applications. Currently, the most popular approach to solve NMF is the multiplicative update algorithm proposed by D.D. Lee and H.S. Seung. In this paper, we propose an additive update algorithm, that has faster computational speed than the algorithm of D.D. Lee and H.S. Seung.**

## I. Introduction

Nonnegative matrix factorization approximation (NMF) is an approximate representation of a given nonnegative matrix $V \in R^{n \times m}$ by a product of two nonnegative matrices $W \in R^{n \times r}$ and $H \in R^{r \times m}$:

$$V \approx W * H \tag{1.1}$$

because $r$ is usually chosen by a very small number, the size of the matrices $W$ and $H$ are much smaller than $V$. If $V$ is a data matrix of some object then $W$ and $H$ can be viewed as an approximate representation of $V$. Thus NMF can be considered an effective technique for representing and reducing data. Although this technique appeared only recently, it has wide application, such as document clustering [7, 11], data mining [8], object recognition [5] and detecting forgery [10, 12]

To measure the approximation in (1.1) often use the Frobenius norm of difference matrix

$$f(W, H) = \frac{1}{2} \|WH - V\|_F^2$$
$$= \frac{1}{2} \sum_{i=1}^{n} \sum_{j=1}^{m} ((WH)_{ij} - V_{ij})^2 \tag{1.2}$$

thus NMF can be formulated as an optimization problem with nonnegative constraints:

$$\min_{W \geq 0, H \geq 0} f(W, H) \tag{1.3}$$

since the objective function is not convex, most methods fail to find the global optimal solution of the problem and only get the stationary point, i.e. the matrix pair *(W, H)* satisfies the Krush-Kuhn-Tucker (KKT) optimal condition [2]:

$$W_{ia} \geq 0, H_{bj} \geq 0$$
$$((WH - V)H^T)_{ia} \geq 0, (W^T(WH - V))_{bj} \geq 0 \tag{1.4}$$
$$W_{ia} * ((WH - V)H^T)_{ia} = 0,$$
$$H_{bj} * (W^T(WH - V))_{bj} = 0, \forall i, a, b, j$$

where

$$(WH - V)H^T = \partial f(W, H) / \partial W$$
$$W^T(WH - V) = \partial f(W, H) / \partial H$$

To update the $H$ or $W$ (the remaining fixed) often use the gradient direction reverse with a certain appropriate steps, so that a reduction in the objective function, on the other still to ensure non-negative of $H$ and $W$. Among the known algorithms solve (1.3) must be mentioned algorithm LS (DD Lee and HS Seung [6]). This algorithm is a simple calculation scheme, easy to install and gives quite good results, so now it remains one of the algorithms are commonly used [10, 12]. LS algorithm is adjusted using the following formula:

$$\tilde{H}_{ij} = H_{ij} - \eta_{ij} \left[ \frac{\partial f}{\partial H} \right]_{ij}$$
$$\equiv H_{ij} + \eta_{ij} (W^T V - W^T WH)_{ij} \tag{1.5}$$

$$\tilde{W}_{ij} = W_{ij} - \varsigma_{ij} \left[ \frac{\partial f}{\partial W} \right]_{ij}$$
$$\equiv W_{ij} + \varsigma_{ij} (V\tilde{H}^T - W\tilde{H}\tilde{H}^T)_{ij} \tag{1.6}$$

by selecting $\eta_{ij}$ and $\varsigma_{ij}$ by the formula:

$$\eta_{ij} = \frac{H_{ij}}{(W^T W H)_{ij}} \ , \ \varsigma_{ij} = \frac{W_{ij}}{(W \widetilde{H} \widetilde{H}^T)_{ij}}$$

then the formula (1.5) and (1.6) become:

$$\widetilde{H}_{ij} = H_{ij} \frac{(W^T V)_{ij}}{(W^T W H)_{ij}} , \widetilde{W}_{ij} = W_{ij} \frac{(V \widetilde{H}^T)_{ij}}{(W \widetilde{H} \widetilde{H}^T)_{ij}}$$

the adjustment formula uses multiplication so this algorithm is called the method of **multiplicative update**. With this adjustment to ensure non-negative of $\widetilde{W}$ and $\widetilde{H}$. In [6] prove the monotonic decrease of the objective function after adjustment:

$$f(\widetilde{W}, \widetilde{H}) < f(W, H)$$

this algorithm LS has the advantages of simple easy to implement on the computer. However, the coefficients $\eta_{ij}$ and $\varsigma_{ij}$ are selected in a special way should not reach the minimum in each adjustment. This limits the speed of convergence of the algorithm.

To improve the convergence speed, E.F. Gonzalez and Y. Zhang has improved LS algorithm by using a coefficient for each column of $H$ and a coefficient for each row of $W$. In other words instead of (1.5) (1.6) using the following formula:

$$\widetilde{H}_{ij} = H_{ij} + \alpha_j \eta_{ij} (W^T V - W^T W H)_{ij}$$
$$\widetilde{W}_{ij} = W_{ij} + \beta_i \varsigma_{ij} (V \widetilde{H}^T - W \widetilde{H} \widetilde{H}^T)_{ij}$$

the coefficients $\alpha_j$ and $\beta_i$ are calculated through the function:

$$\theta = g(A, b, x)$$

(A is the matrix. B and x is the vector). This function is defined as follows:

$$q = A^T (b - Ax) \ \alpha \nu \delta \ \ p = \left[ x./(A^T Ax) \right] \circ q \ ,$$

where the symbol "./" and "$\circ$" denote component-wise division and multiplication, respectively. Then calculate $\theta = g(A, b, x)$ by the formula:

$$\theta = \min \left( \frac{p^T q}{p^T A^T A p}, 0.99 * \max \left\{ \gamma : x + \gamma p \geq 0 \right\} \right)$$

the coefficients $\alpha_j$ and $\beta_i$ are determined by the function $g(A,b,x)$ as follows:

$$\alpha_j = g(W, V_j, H_j), j = 1..n$$
$$\beta_i = g(H^T, V_i^T, W_i^T), i = 1..m$$

However, the experiments showed that improvement of EF Gonzalez and Y. Zhang has not really bring obvious effect. Also as remarks in [3], the LS algorithm and GZ algorithm (EF Gonzalez - Y. Zhang [3]) are not guaranteed the convergence to a stationary point. New algorithm uses addition for updating, so it is called additive update algorithm.

In this paper we propose a new algorithm by updating each element of every matrix $W$ and $H$ based on the idea of nonlinear Gauss - Seidel method [4]. Also with some assumptions, the proposed algorithm ensures reaching stationary point (Theorem 2, subsection III.B). Experiments show that the proposed algorithm converges faster than the algorithms LS and GZ.

The content of the paper is organized as follows. In section 2, we present an algorithm to update an element of the matrix W or H. This algorithm will be used in section 3 to construct a new algorithm for NMF (1.3). We also consider some convergence properties of new algorithm. Section 4 presents a scheme for installing a new algorithm on a computer. In section 5, we present experimental results comparing the calculation speed of algorithms. Finally some conclusions are given in section 6.

## II. ALGORITHM FOR UPDATING AN ELEMENT OF MATRIX

### A. Updating an element of matrix W

In this section, we consider the algorithm for updating an element of $W$, while retaining the remaining elements of $W$ and $H$. Suppose $W_{ij}$ is adjusted by adding $\alpha$ parameter:

$$\widetilde{W}_{ij} = W_{ij} + \alpha \tag{2.1}$$

if $\widetilde{W}$ is an obtained matrix, then by some matrix operations, we have:

$$(\widetilde{W}H)_{ab} = \begin{cases} (WH)_{ab}, a \neq i, b = 1..m \\ (WH)_{ib} + \alpha H_{jb}, a = i, b = 1..m \end{cases}$$

so from (1.2) it follows:

$$f(\widetilde{W}, H) = f(W, H) + g(\alpha) \tag{2.2}$$

where

$$g(\alpha) = \frac{1}{2}(p\alpha^2) + q\alpha \qquad (2.3)$$

$$p = \sum_{b=1}^{m} H_{jb}^2 \qquad (2.4)$$

$$q = \sum_{b=1}^{m} (WH - V)_{ib} * H_{jb} \qquad (2.5)$$

to minimize $f(\widetilde{W}, H)$, one needs to define $\alpha$ so that $g(\alpha)$ achieves the minimum value on the condition $\widetilde{W}_{ij} = W_{ij} + \alpha \geq 0$. Because $g(\alpha)$ is a quadratic function, then $\alpha$ can be defined as follows:

$$\alpha = \begin{cases} 0, q = 0 \\ \max(\dfrac{-q}{p}, -w_{ij}), q > 0 \\ \dfrac{-q}{p}, q < 0 \end{cases} \qquad (2.6)$$

formula (2.6) always means because if $q \neq 0$ then by (2.4) we have $p>0$

From (2.3) and (2.6), we get:

$g(\alpha) = 0$, if (q = 0) or (q>0 and $W_{ij}$ =0) (2.7.a)

$g(\alpha) < 0$, otherwise (2.7.b)

By using update formulas (2.1) and (2.6), the monotonous decrease of the objective function *f(W,H)* is confirmed in the following lemma .

*LEMMA 1: If conditions KTT are not satisfied at $W_{ij}$, then:*

$$f(\widetilde{W}, H) < f(W, H)$$

*Otherwise:* $\widetilde{W} = W$

*Proof.* From (2.4), (2.5) it follows

$$q = ((WH - V)H^T)_{ij}$$

Therefore, if conditions KTT (1.4) are not satisfied at $W_{ij}$, then properties

$$W_{ij} \geq 0, q \geq 0, W_{ij} * q = 0 \qquad (2.8)$$

cannot occur simultaneously. From this and because $W_{ij} \geq 0$, it follows that case (2.7.a) cannot happen. So case (2.7.b) must occur and we have $g(\alpha) < 0$. Therefore, from (2.2) we obtain

$$f(\widetilde{W}, H) < f(W, H)$$

Conversely, if (2.8) is satisfied, it means that: *q=0* or *q>0* and $W_{ij} = 0$. So from (2.6), it follows $\alpha = 0$. Therefore, by (2.1) we have

$$\widetilde{W}_{ij} = W_{ij}$$

Thus lemma is proved.

*B. Updating an element of matrix H*

Let $\widetilde{H}$ be matrix obtained from the update rule:

$$\widetilde{H}_{ij} = H_{ij} + \beta \qquad (2.9)$$

where $\beta$ is defined by the formulas:

$$u = \sum_{a=1}^{n} W_{ai}^2 \qquad (2.10)$$

$$v = \sum_{a=1}^{n} W_{ai} * (WH - V)_{aj} \qquad (2.11)$$

$$\beta = \begin{cases} 0, v = 0 \\ \max(\dfrac{-v}{u}, -H_{ij}), v > 0 \\ \dfrac{-v}{u}, v < 0 \end{cases} \qquad (2.12)$$

By the same discussions used in lemma 1, we have

*LEMMA 2: If conditions KTT are not satisfied at $H_{ij}$, then:*

$$f(W, \widetilde{H}) < f(W, H)$$

*Otherwise:* $\widetilde{H} = H$

III. THE PROPOSED ALGORITHM

*A. Updating matrices W and H*

In this section we consider the transformation *T* from *(W, H)* to $(\widetilde{W}, \widetilde{H})$ as follows:

- Modify elements of *W* by subsection II.A

- Modify elements of $H$ by subsection II.B

In other words, the transformation $(\widetilde{W}, \widetilde{H}) = T(W, H)$ shall be carried out as follows:

Step 1: Initialise

$$\widetilde{W} = W, \widetilde{H} = H$$

Step 2: Update elements of $\widetilde{W}$

For $j=1,...,r$ and $i=1,...,n$

$$\widetilde{W}_{ij} \leftarrow \widetilde{W}_{ij} + \alpha$$

$\alpha$ is computed by (2.4)-(2.6)

Step 3 : Update elements of $\widetilde{H}$

For $i=1,...,r$ and $j=1,...,m$

$$\widetilde{H}_{ij} \leftarrow \widetilde{H}_{ij} + \beta$$

$\beta$ is computed by (2.10) - (2.12)

From Lemmas 1 and 2, we easily obtain the following important property of the transformation $T$.

*LEMMA 3: If solution (W,H) does not satisfy the condition KTT (1.4), then*

$$f(\widetilde{W}, \widetilde{H}) = f(T(W,H)) < f(W,H)$$

*In the contrary case, then:* $(\widetilde{W}, \widetilde{H}) = (W, H)$

Following property is directly obtained from Lemma 3.

*COROLLARY 1: For any* $(W,H) \geq 0$, *if set*

$$(\widetilde{W}, \widetilde{H}) = T(W, H)$$

*then* $(\widetilde{W}, \widetilde{H}) = (W, H)$ *or* $f(\widetilde{W}, \widetilde{H}) < f(W, H)$

**B. Algorithm for NMF (1.3)**

The algorithm is described through the transformation $T$ as follows:

Step1. Initialize $W=W^1>=0$, $H=H^1>=0$

Step 2. For $k=1,2,...$

$$\left(W^{k+1}, H^{k+1}\right) = T\left(W^k, H^k\right)$$

From Corollary 1, we obtain the following important property of above algorithm.

*THEOREM 1. Suppose* $(W^k, H^k)$ *is a sequence of solutions created by algorithm 3.2, then the sequence of*

objective function values $f(W^k, H^k)$ *actually decreases monotonously:*

$$f(W^{k+1}, H^{k+1}) < f(W^k, H^k), \forall k \geq 1$$

Moreover, the sequence $f(W^k, H^k)$ is bounded below by zero, so Theorem 1 implies the following corollary.

*COROLARRY 2. Sequence* $f(W^k, H^k)$ *is a convergence sequence. In other words, there exists non-negative value f such that:*

$$\lim_{k \to \infty} f(W^k, H^k) = \bar{f}$$

Now we consider another convergence property of Algorithm III.B.

*THEOREM 2. Suppose* $(\overline{W}, \overline{H})$ *is a limit point of the sequence* $(W^k, H^k)$ *and*

$$\sum_{b=1}^{m} \overline{H}_{jb}^2 > 0, j = 1...r \qquad (3.1)$$

$$\sum_{a=1}^{n} \overline{W}_{ai}^2 > 0, i = 1...r \qquad (3.2)$$

*Then* $(\overline{W}, \overline{H})$ *is the stationary point of the problem (1.3)*

Proof. By assumption, $(\overline{W}, \overline{H})$ is the limit of some subsequence $(W^{t_k}, H^{t_k})$ of the sequence $(W^k, H^k)$:

$$\lim_{k \to \infty} (W^{t_k}, H^{t_k}) = (\overline{W}, \overline{H}) \qquad (3.3)$$

By conditions (3.1), (3.2), the transformation T is continuous at $(\overline{W}, \overline{H})$. Therefore from (3.3) we get:

$$\lim_{k \to \infty} T(W^{t_k}, H^{t_k}) = T(\overline{W}, \overline{H})$$

Moreover, since $T(W^{t_k}, H^{t_k}) = (W^{t_k+1}, H^{t_k+1})$, then:

$$\lim_{k \to \infty} (W^{t_k+1}, H^{t_k+1}) = T(\overline{W}, \overline{H}) \qquad (3.4)$$

Using a continuation of the object function f(W,H), from (3.3), (3.4) we have

$$\lim_{k \to \infty} f(W^{t_k}, H^{t_k}) = f(\overline{W}, \overline{H})$$

$$\lim_{k \to \infty} f(W^{t_k+1}, H^{t_k+1}) = f(T(\overline{W}, \overline{H}))$$

Because, on the other hand, by Corollary 2, sequence $f(W^k, H^k)$ is convergent, it follows:

$$f(T(\overline{W}, \overline{H})) = f(\overline{W}, \overline{H})$$

Therefore, by Lemma 3, $(\overline{W}, \overline{H})$ must be a stationary point of problem (1.3).

Thus theorem is proved.

### IV. SOME VARIATIONS OF ALGORITHM

In this section we provide some variations for algorithm in subsection III.B (Algorithm III.B) to reduce the volume of calculations and increase convenience for the installation program.

#### A. Evaluate computational complexity

To update an element $W_{ij}$ by formulas (2.1), (2.4), (2.5), (2.6), we need to use m multiplications for calculating $p$ and $m*(n*m*r)$ multiplications for calculating $q$. Similarly to update element $H_{ij}$ by using (2.9), (2.10), (2.11), (2.12), we need n multiplications for computing $u$ and $(n*m*r)*n$ multiplications for computing $v$. It follows that the number of calculations to make a loop ($transformation(\widetilde{W}, \widetilde{H}) = T(W, H)$) of the Algorithm III.B is

$$2*n*m*r*(1 + n*m*r) \tag{4.1}$$

#### B. Some variations for updating W and H

##### 1) Updatting Wij

If set

$$D = WH - V \tag{4.2}$$

then the formula (2.5) for q becomes:

$$q = \sum_{b=1}^{m} D_{ib} * H_{jb} \tag{4.3}$$

if one considers D as known, the calculation of q in (4.3) needs m multiplications. After updating $W_{ij}$ by the formula (2.1), we need to recalculate the D from $\widetilde{W}$ to be used for the adjustment of other elements of W:

$$\widetilde{D} = \widetilde{W}\widetilde{H} - V$$

from (2.1) and (4.2), it is seen that $\widetilde{D}$ is determined from D by the formula:

$$\widetilde{D}_{ab} = \begin{cases} D_{ab}, a \neq i, b = 1..m \\ D_{ib} + \alpha H_{jb}, a = i, b = 1..m \end{cases} \tag{4.4}$$

So we only need to adjust the i$^{th}$ row of D and need to use m multiplications.

From formulas (2.1), (2.4), (2.6), (4.3) and (4.4), we have a new scheme for updating matrix W as follows.

##### 2) Scheme for updating matrix W

For j = 1 To r

$$p = \sum_{b=1}^{m} H_{jb}^2$$

For i=1 To n

$$q = \sum_{b=1}^{m} D_{ib} * H_{jb}$$

$$\alpha = \begin{cases} 0, q = 0 \\ \max(\dfrac{-q}{p}, -w_{ij}), q \neq 0 \end{cases}$$

$$W_{ij} \leftarrow W_{ij} + \alpha$$

End For i

End For j

The total number of operations used to adjust the matrix W is: $2 \times n \times m \times m \times r + m \times r$

##### 3) Updating Hij

Similarly, the formula (2.11) for v becomes

$$v = \sum_{a=1}^{n} W_{ai} * D_{aj} \tag{4.5}$$

according to this formula, we only use n multiplications to calculate $v$. After adjusting $H_{ij}$ by formula (2.9), we need to recalculate matrix D by the following formula:

$$\widetilde{D}_{ab} = \begin{cases} D_{ab}, b \neq j, a = 1..n \\ D_{aj} + \beta H_{ai}, b = j, a = 1..n \end{cases} \tag{4.6}$$

So we only need to adjust the j$^{th}$ column of D and need to use n multiplications.

From formulas (2.9), (2.10), (2.12), (4.5) and (4.6), we have a new scheme for updating matrix H as follows.

##### 4) Scheme for updating matrix H

For i = 1 To r

$$u = \sum_{a=1}^{n} W_{ai}^2$$

For j=1 To m

$$v = \sum_{a=1}^{n} W_{ai} * D_{aj}$$

$$\beta = \begin{cases} 0, v = 0 \\ \max(\dfrac{-v}{u}, -H_{ij}), v \neq 0 \end{cases}$$

$$H_{ij} \leftarrow H_{ij} + \beta$$

$$D_{ib} \leftarrow D_{ib} + \beta H_{jb}, a = i, b = 1..m$$

End for j

End for i

The total number of operations used to adjust the matrix $H$ is: $2 \times n \times m \times r + n \times r$.

Using the above results together, we can construct a new calculating scheme for the Algorithm III.B as follows.

*C. New calculating scheme for the Algorithm III.B*

1. Initialize W=W$^1$>=0, H=H$^1$>=0

        D=WH-V

2. For *k=1,2,...*

- Update W by using subsection IV.B.2
- Update H by using subsection IV.B.4

the computational complexity of this scheme is as follows:

- Initialization step needs *n\*m\*r* multiplications for computing *D*.
- Each loop needs *n\*m\*r + r\*(n+m)* multiplications.

Comparing with (4.1), number of operations has now greatly reduced.

## V. EXPERIMENTS

In this section, we present results of 2 experiments on the algorithms: New NMF (new proposed additive update algorithm), GZ and LS. The programs are written in MATLAB and run on a machine with configurations: *Intel Pentium Core 2 P6100 2.0 GHz, RAM 3GB*. New NMF is built according to the schema in subsection IV.C.

*A. Experiment 1*

Used to compare the speed of convergence to stationary point of the algorithms. First of all condition KKT (1.4) is equivalent to the following condition:

$$\delta(W, H) = 0$$

where:

$$\delta(W, H) = \sum_{i=1}^{n} \sum_{a=1}^{r} \left| \min(W_{ia}, ((WH - V)H^T)_{ia}) \right|$$

$$+ \sum_{b=1}^{r} \sum_{j=1}^{m} \left| \min(H_{bj}, (W^T(WH - V)_{bj}) \right|$$

Thus if *h(W, H)* is smaller, then *(W,H)* is closer to the stationary point of the problem (1.3). To get a quantity independent with the size of *W* and *H*, we use following formula:

$$\Delta(W, H) = \frac{\delta(W, H)}{\delta_W + \delta_H}$$

in which $\delta_W$ is the number of elements of the set:

$$\left\{ \min(W_{ia}, ((WH - V)H^T)_{ia}) \neq 0 \mid i = 1...n, a = 1...r \right\}$$

and $\delta_H$ is the number of elements of the set:

$$\left\{ \min(H_{bj}, (W^T(WH - V)_{bj}) \neq 0 \mid j = 1...m, b = 1...r \right\}$$

$\Delta(W, H)$ is called a normalized KKT residual. Table 1 presents the value $\Delta(W, H)$ of the solution *(W, H)* received by each algorithm implemented in given time periods on the data set of size (n, m, r) = (200,100,10) in which *V, W$^1$, H$^1$* was generated randomly with $V_{ij} \in [0,500]$, $(W^1)_{ij} \in [0,5]$, $(H^1)_{ij} \in [0,5]$.

TABLE I.     NORMALIZED KKT RESIDUALS VALUE $\Delta(W, H)$

| Time (sec) | New NMF | GZ | LS |
|---|---|---|---|
| 60 | 3.6450 | 3700.4892 | 3576.0937 |
| 120 | 1.5523 | 3718.2967 | 3539.8986 |
| 180 | 0.1514 | 3708.6043 | 3534.6358 |
| 240 | 0.0260 | 3706.4059 | 3524.6715 |
| 300 | 0.0029 | 3696.7690 | 3508.3239 |

The results in Table 1 show that the two algorithms GZ and LS cannot converge to a stationary point (value $\Delta(W, H)$ is still large). Meanwhile, New NMF algorithm still possible converges to a stationary point because value $\Delta(W, H)$ reaches of approximately equal value *0*.

*B. Experiment 2*

Used to compare the convergence speed to the minimum value of objective function *f(W, H)* of the algorithms implemented in given time periods on the data set of size *(n, m, r) = (500,100,20)*, in which *V, W$^1$, H$^1$* was generated randomly with $V_{ij} \in [0,500]$, $(W^1)_{ij} \in [0,1]$, $(H^1)_{ij} \in [0,1]$. The algorithms are run 5 times with 5 different pairs of $W^l$, $H^l$ generated randomly in the interval $[0,1]$. Average values of objective function after 5 times of performing the algorithms in each given time period are presented in Table 2.

TABLE II.    AVERAGE VALUES OF OBJECTIVE FUNCTION

| Time (sec) | New NMF | GZ | LS |
|---|---|---|---|
| 60 | 57.054 | 359.128 | 285.011 |
| 120 | 21.896 | 319.674 | 273.564 |
| 180 | 18.116 | 299.812 | 267.631 |
| 240 | 17.220 | 290.789 | 264.632 |
| 300 | 16.684 | 284.866 | 262.865 |
| 360 | 16.458 | 281.511 | 261.914 |

The results in Table 2 show that the objective function value of the solutions generated by two algorithms GZ and LS is quite large. Meanwhile the objective function value of New NMF algorithm is much smaller.

## VI. CONCLUSIONS

This paper proposed a new additive update algorithm for solving the problem of nonnegative matrix factorization. Experiments show that the proposed algorithm converges faster than the algorithms LS and GZ. The proposed algorithm has a simple calculation scheme too, so it is easy to install and use in applications.

### REFERNCES